\documentclass[aps,print,pre,lineo,11pt]{revtex4-2}
\usepackage{lipsum}
\usepackage{amsmath}
\usepackage{graphicx}
\usepackage{dcolumn}
\usepackage{bm}
\usepackage[mathlines]{lineno}
\usepackage{setspace}
\usepackage{color}
\usepackage[utf8]{inputenc}
\usepackage[T1]{fontenc}
\usepackage{etoolbox}
\usepackage{hyperref}
\usepackage{comment}
\hypersetup{
    colorlinks=true,
    linkcolor=black,
    filecolor=black,      
    urlcolor=black,
    citecolor=black,
    }
\usepackage[capitalise]{cleveref}
\begin{document}

\title{Heavy Particle Motion in Rotational Vortices}
\author{Orr Avni}
\affiliation{School of Engineering, Brown University, 184 Hope St., Providence, Rhode Island 02912, United States}
\author{Alok Kumar}
\affiliation{Faculty of Aerospace Engineering, Technion - Israel Institute of Technology, Haifa, 320003, Israel.}
\author{Yuval Dagan}
\email{yuvalda@technion.ac.il}
\affiliation{Faculty of Aerospace Engineering, Technion - Israel Institute of Technology, Haifa, 320003, Israel.}
\date{\today}

\begin{abstract}

This study examines the motion of spherical inertial particles in a three-dimensional rotating cylindrical vortex -- a simplified model of geophysical flow structures such as oceanic eddies.
The analytical vortex formulation enables the isolation of the key mechanisms that govern particle transport in rotating flows.
Using Lagrangian particle tracking simulations, we investigate the influence of drag, buoyancy, virtual mass, Coriolis, and Magnus lift forces across a range of particle sizes, densities, and vortex rotation rates. Results show that particle aggregation and periodic stability depend on both particle inertia and flow parameters. Rotational lift forces, though often neglected for spherical particles, become dominant at moderate particle Stokes numbers and introduce slow vertical oscillations in both particle position and spin. The balance of forces determines whether particles settle into stable periodic orbits or escape the vortex. 
Our analysis reveals unique equilibrium positions that exhibit bifurcations, with multiple or vanishing steady states depending on particle and flow characteristics. 
Our results demonstrate how particle aggregation and orbital stability stem from the complex coupling between rotational lift, Coriolis, drag, buoyancy, and virtual mass forces.
This model may promote informed modeling of dispersed phase transport in marine flows and industrial mixing processes.

\end{abstract}
\maketitle
\newpage

\section{Introduction}\label{sec:intro}

Rotating fluid structures are ubiquitous in nature and engineering, spanning a range of scales: from astrophysical phenomena, through planetary atmospheres to small-scale vortical structures in turbulence \citep{ogilvie_astrophysical_2016,biferale2016coherent}.  
Rotating flows are also important in industrial contexts, such as reactor chambers, mixers, and crystal growth applications \citep{greenspan1990,hopfinger1993,dold_rotating_1999}.
Given the current challenge of increased oceanic microplastic pollution, a detailed investigation of the multiphase behavior of coherent vortical structures in the ocean, such as mesoscale eddies and submesoscale fronts, becomes urgent \citep{andrady_microplastics_2011,sutherland2023}.
These rotating flows govern large-scale transport of heat, salinity, nutrients, and pollutants, all of which are central to ocean health and climate forecasting \citep{largescaledudley,onink_role_2019,rahmani_aggregation_2022}. 

Despite that, accurately modeling particle motion in vortical structures presents substantial challenges \citep{Balachandar2010,Voth2017,brandt_particle-laden_2022}. 
These include the wide range of spatial and temporal scales involved, the nonlinear interactions between the carrier flow and particle inertia, and the computational cost of fully resolved multiphase simulations \citep{beron-vera_nonlinear_2021}.
Predicting the fate of microplastics, for instance, requires capturing interactions between particles and coherent flow structures across scales and regimes \citep{sutherland2023}.
Even simplified numerical approaches, such as high-fidelity Lagrangian particle tracking, can become computationally expensive for realistic oceanic conditions.
On the other hand, using simplified theoretical flow models, one may employ Lagrangian methods to track the particle and droplet trajectories and estimate their dispersion~\cite{Marcu1996I,Marcu1996II, Marcu1996mixing,Dagan2021,Avni2022a} and entrapment within coherent vortical structures~\cite{Marcu1995burgers,IJzermans2006,Angilella2010, Ravichandran2022,avni_droplet_2023,avni_droplet_2023II}.
Taking such an approach allows the isolation of specific transport mechanisms, including Brownian motion~\cite{wang_brownian_2024}, oscillatory flows~\cite{Dagan2017b,dagan2017particle,daganSimilarityFlames2018}, aerosol formation~\cite{Avni2022b}, and complex formations~\cite{Yerasi2022}, while studying their influence on the particle dynamics.

In the context of oceanic transport, simplified wave-current flow models were used to study the dispersion of inertial, non-spherical particles \cite{dibenedetto_transport_2018,clark_dispersion_2023,sunberg_parametric_2024}, where the complex nature and the added degrees of freedom call for the use of simplified flow models.
A combination of direct numerical simulations and simplified models was employed by \citet{rypina2015c,pratt2014,rypina2024}, who investigated inertial particle motion in large-scale vortices using an analytically defined axisymmetric flow field derived from the incompressible Navier–Stokes equations in a rotating cylindrical domain.
This model has proven effective in capturing the essential structure and dynamics of coherent oceanic vortices while avoiding the complexity of numerical simulations.
The model reproduced key features of the DNS solution and has been used to study slightly buoyant particles in quasi-chaotic regimes \cite{rypina2024}.

This study revisits the model presented by \citet{rypina2024} while adopting a different approach: we retain the symmetric, steady structure of the background flow while extending the investigation to particles that are not necessarily near-buoyant or small.
Allowing for a range of density ratios between the carrier fluid and the dispersed phase, we may address a broader class of particles, such as sediment grains, organic aggregates, and synthetic debris.
However, we neglect the effects of externally imposed perturbations that induce chaotic trajectories; instead, we focus on the unsteady behavior induced by the dispersed phase nonlinear dynamics in laminar rotating vortices. 
This extension necessitates incorporating additional effects that are often neglected when addressing the dispersion of spherical particles; chief among them is their rotational dynamics.
Even for spherical particles, rotational slip relative to the surrounding flow induces lift forces that may affect particle trajectories at finite Stokes numbers.
The inclusion of such effects enables a more complete picture of inertial particle motion in rotation-dominated flows.
Our goals are twofold: first, to characterize the coupled translational and rotational dynamics governing particle dynamics, and second, to identify the mechanisms that govern the system's dynamic response.
To that end, a parametric investigation of the particle size, density ratio, and vortex circulation influence is employed to map the mechanisms driving aggregation and dispersion of particles in large vortical structures, serving as a reduced-order representation of the interaction between dispersed phases and oceanic vortices.

The remainder of the paper is organized as follows. In \cref{sec:gov}, we present the analytical vortex model used in our study (\cref{sec:flow}) and particle equations of motion mathematical formulation (\cref{sec:lagra}). 
\Cref{sec:dynamics} analyzes the particle dynamics under varying physical parameters, focusing on the roles of inertia, density ratio, and rotational forces. 
\Cref{sec:ss} presents the equilibrium state of particles and analyzes its sensitivity to changes in vortex and particle properties. 
Finally, conclusions and perspectives for future work are provided in Section~\ref{sec:conc}.

\section{Governing Equations}\label{sec:gov}
We analyze the transport of solid, perfectly spherical particles within a three-dimensional, steady, analytically described rotational vortex.
Using a Lagrangian framework, we track the spatial location $\bm{x}$, velocity $\bm{v}$, and angular velocity $\bm{\omega}$ of particles and couple them to the local flow field $\bm{u}$.
We assume that the particles are sufficiently dilute such that particle-particle interactions are negligible and the motion of particles within the fluid does not alter the carrier flow.
The governing equations for the carrier flow and the Lagrangian particles are presented as follows.

\subsection{Carrier flow}\label{sec:flow}
The carrier flow is modeled as a three-dimensional, eddy-like rotating flow characterized by both vertical and horizontal circulation.
This phenomenological flow field is derived from incompressible motion in a rotating cylindrical domain \citep{pratt2014,hopfinger1993,greenspan1990}; the model has been studied and applied in the context of oceanic circulation \citep{pratt2014,rypina2015c,rypina2024}.
Although this model simplifies the representation of large turbulent structures, it captures the essential dynamics of rotational vortices.
The analytic model reproduces many of the qualitative features of the numerically obtained velocity field in rotational cylindrical flows and was found to effectively describe particle behavior in such conditions \citep{rypina2024}.

Assuming a cylinder of height $h$ and aspect ratio $a$ is rotating along its vertical axis at a frequency $\phi'_z$, the analytical velocity field of the rotation-induced flow within the cylinder yields
\begin{equation} \label{eq:flow}
        \bm{u} = 2\left(1-2z\right)\frac{r(r-a)}{a}\bar{e}_r + \phi_zr\bar{e}_\theta +2z(1-z)\frac{2a-3r}{a}\bar{e}_z
\end{equation}
in a stationary frame of reference.
The velocity and length are scaled by $U$, the maximal vertical circulatory velocity, and $h$, the vortex depth, correspondingly, while $\phi_z = \phi'_z h /U$ is the normalized rotational frequency.
The flow is defined only within the cylinder: that is, $0<z<1$ and $0<r<a$ so that fluid particles cannot escape its domain.

\subsection{Particulate matter}\label{sec:lagra}
\citet{Maxey1983} formulated generalized equations of motion for small particles in nonuniform, unsteady flows, considering both gravity, drag, virtual mass, and the Basset "history" force.
The particle's characteristic length is significantly smaller than the steady vortex's size, which allows for neglecting the influences of Faxen's drag correction and the terms related to the particle's history.
On the other hand, given the extended range of particle properties, we argue that incorporating rotational inertia effects, i.e., spin-induced lift forces, is crucial in order to capture the particle dynamics under different flow conditions.
Hence, we may now reduce the generalized form of the dimensionless momentum equation to
\begin{equation}\label{eq:MR_ND}
\left( 1 + \frac{1}{2} \overline{\rho} \right) \frac{d\tilde{\bm{v}}}{dt} =\text{St}^{-1} (\tilde{\bm{u}} - \tilde{\bm{v}}) + \frac{3}{2} \overline{\rho} \frac{D\tilde{\bm{u}}}{Dt} + (1 - \overline{\rho}) \text{Fr}^{-2} + \frac{3}{4} \overline{\rho} \left[\Omega\times({\tilde{\bm{u}}}- \tilde{\bm{v}}) \right],
\end{equation}
where $\text{St} = \frac{U\rho_p D_p^2}{18h\mu_f}$ is the particle's Stokes number, $\text{Fr} = \frac{U}{\sqrt{gh}}$ is the Froud number, $\bar{\rho}=\frac{\rho_f}{\rho_p}$ is the density ratio between the particle and the medium, and
\begin{equation}
    \Omega = \frac{1}{2}\nabla\times\bm{u} - \bm{\omega} 
\end{equation}
is the sphere's relative rate of rotation with respect to the fluid.
We introduce the rotational virtual forces by transforming into a stationary frame of reference \citep{beron-vera2019, rypina2024}
\begin{equation}
    \begin{gathered}
    \tilde{\bm{u}} = \bm{u} - \phi_z\bar{e}_z  \times \bm{x}\\
    \frac{D\tilde{\bm{u}}}{Dt} = \frac{D\bm{u}}{Dt} - \frac{d}{dt}\left( \phi_z\bar{e}_z  \times \bm{x}\right) + \phi_z\bar{e}_z  \times \phi_z\bar{e}_z  \times \bm{x}.
    \end{gathered}
\end{equation}
Substituting and reordering, we obtain three nonlinear second-order ordinary differential equations (ODEs):
\begin{equation}\label{eq:mom}
    \begin{gathered}
    \frac{dv_r}{dt} = \frac{2}{(2+\rho)\text{St}}(u_r - v_r) + \frac{3\rho}{(2+\rho)}\left[\left(\bm{u}\cdot\nabla\right)\bm{u}\right]_{r}+ \omega^2_0\left(\frac{2\rho - 2}{2+\rho}\right) + 2\phi_z\left(v_\theta - \frac{3\rho}{2+\rho}u_\theta\right) +\\ +\frac{3\rho}{2(2+\rho)}\left[(\phi_\theta-\omega_\theta)(u_z - v_z) - (\phi_z-\omega_z)(u_\theta - v_\theta)\right], \\
   \frac{dv_\theta}{dt} = \frac{2}{(2+\rho)\text{St}}(u_\theta - v_\theta)  - 2\phi_z\left(v_r - \frac{3\rho}{2+\rho}u_r\right)+\\ 
    +\frac{3\rho}{2(2+\rho)}\left[(\phi_z-\omega_z)(u_r - v_r) + \omega_r(u_z - v_z)\right], \\
    \frac{dv_z}{dt} = \frac{2}{(2+\rho)\text{St}}(u_z - v_z) + \frac{3\rho}{(2+\rho)}\left[\left(\bm{u}\cdot\nabla\right)\bm{u}\right]_{z} - \frac{2(1-\rho)}{(2+\rho)\text{Fr}^2} -\\ -\frac{3\rho}{2(2+\rho)}\left[\omega_r(u_\theta - v_\theta) + (\phi_\theta-\omega_\theta) (u_z - v_z)\right].
    \end{gathered}
\end{equation}
To resolve the Magnus forces, we incorporate the particle's rotational dynamic equation \citep{Crowe2011}
\begin{equation}\label{eq:omega_D}
    \frac{d(I_p\bm{\omega})}{dt} = T_f - \phi_z\bar{e}_z\times(I_p\bm{\omega}),
\end{equation}
where $I_p$ is the particle's moment of inertia, and $T_f$ is the flow-induced torque acting on the particle. 
Assuming particle sphericity, the dimensionless form of \cref{eq:omega_D} given as a set of ODEs is
\begin{equation}\label{eq:spin}
\begin{gathered}
    \frac{d\omega_r}{dt} = -\frac{10}{3\text{St}} \omega_r + \phi_z\omega_\theta,\\
    \frac{d\omega_{\theta}}{dt} = \frac{10}{3\text{St}}(\phi_\theta - \omega_\theta) - \phi_z\omega_r,\\
    \frac{d\omega_z}{dt} = \frac{10}{3\text{St}}(\phi_z - \omega_z).
\end{gathered}
\end{equation}

Hence, we may model the complete dynamic behavior of a Lagrangian particle within a rotating vortex in terms of nine state variables $X_p = \left[\bm{x},\bm{v},\bm{\omega}\right]$.
The complex coupling between the external, vortex-induced forcing and the particle's response, \cref{eq:mom,eq:spin} gives rise to a nonlinear dynamic system, $\dot{\bm{X}}_p = F(\bm{X}_p,\dot{\bm{X}}_p)$.
We proceed to analyze the system transient response by numerically resolving the ODEs for $X_{p}(t)$ in the following section.

\section{Particle dynamics}\label{sec:dynamics}
The coupled initial value problem governing the system dynamics is resolved using a fourth-order Runge–Kutta numerical scheme. 
Unless otherwise specified, particles are initially placed at the flow's stagnation point $(r, z) = \left( \frac{2}{3}a, \frac{1}{2} \right)$; due to the spherical symmetry of the vortex, the initial azimuthal coordinate is arbitrary and does not influence the physical outcome.
The particles' initial velocity and spin are set to match the local flow velocity and vorticity, respectively.
Here, we assume the vortex geometry remains constant and set a depth-to-radius ratio of $a = \frac{1}{2}$.
Correspondingly, we fix the Froude number at $\text{Fr}^{-2} = 10$, corresponding to a vortex depth and vertical velocity in the order of unity.
Finally, the physicality of the analysis is maintained by limiting the computational domain to the cylindrical vortex domain; particles crossing out of it are eliminated from the simulation.

To investigate the particle dynamics across various regimes, we vary the Stokes number from $\text{St} = 0.001$ to $1$.
The fluid-to-particle density ratio, $\bar{\rho}$, which directly influences the centrifugal, Coriolis, buoyancy, and Magnus forces, is varied between $\bar{\rho}=0.01$ and $\bar{\rho}=1.05$.
The vortex rotational frequency $\phi_z$ is studied in the range $\left[0.02\pi,2\pi\right]$ to examine how variations in the ratio of vertical recirculation time to azimuthal circulation time affect the particle motion.
Unless otherwise indicated, the default parameter values are $\bar{\rho} = 1$ and $\phi_z = 2\pi\times1$; i.e., neutrally buoyant particles rotating within a vortex of equal characteristic timescales for both the vertical and azimuthal recirculation.

\begin{figure}
    \centering
    \includegraphics[width=1\linewidth]{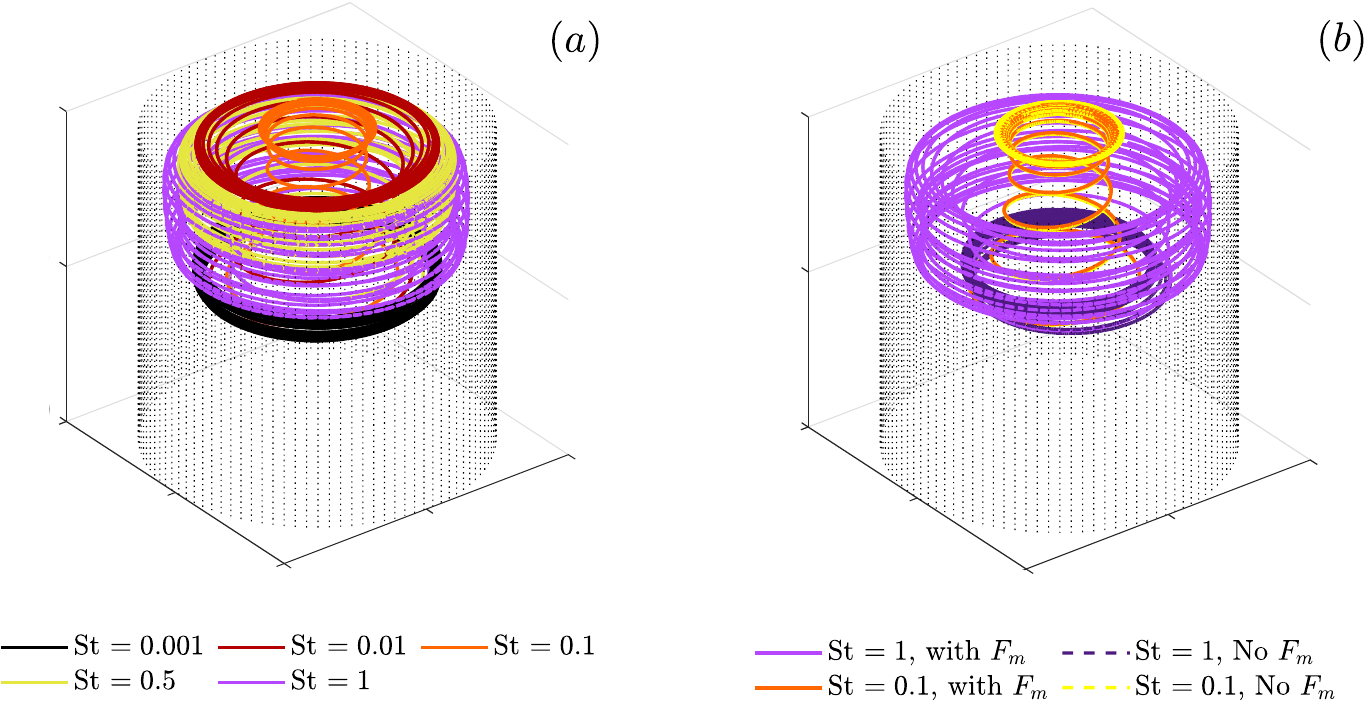}
    \caption{Three-dimensional trajectories of neutrally buoyant $\bar{\rho}=1$ spherical particles within a vortex rotating at a frequency $\phi_z = 2\pi$. (a) Particle trajectories for different Stokes numbers $\text{St}$. (b) Comparison between the obtained trajectories with Magnus lift forces (with $F_m$, solid lines) and without (no $F_m$, dashed lines) for two Stokes values $\text{St}=0.1,1$.}
    \label{fig:1}
\end{figure}

\Cref{fig:1} showcases the three-dimensional trajectories of particles with varying Stokes numbers and demonstrates the Magnus lift force's influence through comparative analysis of trajectories with and without this force.
As shown in \cref{fig:1}(a), low Stokes particles $\text{St} \leq 0.001$ maintain stable orbits near the flow stagnation point, which represents the steady orbit of a fluid particle.
However, even at $\text{St} = 0.001$, the particles deviate from perfect tracer behavior due to drag forces, resulting in a slight upward shift in the stable orbit vertical position.
This upward displacement intensifies at $\text{St}=0.01$, where particles migrate from the stagnation point and establish stable orbits in proximity to the vortex's upper boundary.
For $\text{St} = 0.1$, the stable orbit maintains its vertical position while significantly reducing its diameter.
As the Stokes number increases beyond $0.1$, the trend reverses as the orbits expand radially, significant vertical oscillations emerge, and the particles do not stabilize around a clear steady-state periodic trajectory.
\Cref{fig:1}(b) reveals the Magnus lift force's critical role at large Stokes numbers.
For $\text{St} \leq 0.1$, the Magnus force exerts minimal influence on the particle dynamics, evidenced by the convergence of solid and dashed trajectory curves.
However, at $\text{St} = 1$, the Magnus force substantially alters the particle motion, generating significant upward translation and vertical oscillations.

\begin{figure}
    \centering
    \includegraphics[width=1\linewidth]{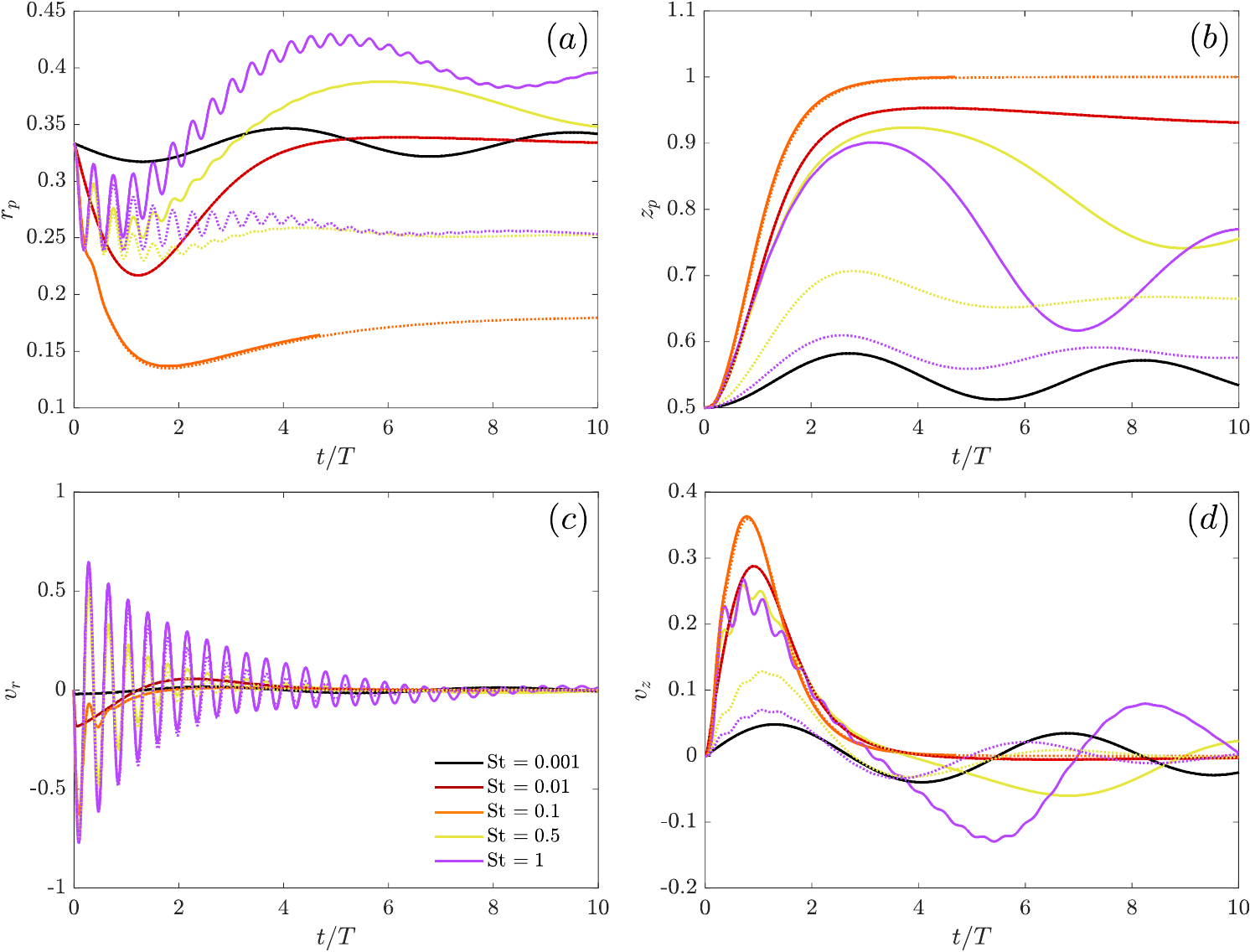}
    \caption{Dynamic response of neutrally buoyant $\bar{\rho}=1$ spherical particles within a vortex rotating at a frequency $\phi_z = 2\pi$ for particle Stokes number in the range $\text{St}=0.001 - 1$. Times are normalized by the vortex rotational period $T=\phi_z/2\pi$. (a)  Particle Radial location $r_p(t)$. (b) Particle axial location $z_p(t)$. (c) Particle radial velocity $v_r(t)$. (d) Particle axial velocity  $v_z(t)$.}
    \label{fig:2}
\end{figure}

\Cref{fig:2} demonstrates the transient response of neutrally buoyant spherical particles corresponding to those in \cref{fig:1}.
At the lowest Stokes number $\text{St} = 0.001$, particles exhibit sustained low-frequency oscillatory motion in both the radial and vertical directions.
These oscillations arise from the combined influence of Coriolis forces and virtual mass effects, leading to a deviation from tracer-like behavior despite the low Stokes value.
For intermediate Stokes numbers $0.01 \leq \text{St} \leq 0.1$, drag effects lead to complete damping of the oscillations, and the particles gradually settle into steady-state trajectories following their departure from the flow stagnation point.
As evident in \cref{fig:2}(c), high-frequency radial oscillations emerge for $\text{St} \geq 0.5$ due to increased particle inertia and the resulting lengthening of the particle relaxation time.
These oscillations induce secondary vertical oscillations due to the coupled nature of the system; however, both dissipate completely within approximately five to ten turnover times.
Additionally, low-frequency oscillations in the vertical position of the particle appear at this Stokes range. 
However, these low-frequency oscillations diminish in particles simulated without the Magnus force, suggesting they are driven by rotational lift forces acting on the particle. 
The nature of the two oscillatory modes differs fundamentally: the rapid oscillations are governed by the combined effects of inertia, Coriolis, and virtual mass forces, while low-frequency oscillations in the vertical direction primarily result from the interaction of inertia and Magnus forces.

\begin{figure}
    \centering
    \includegraphics[width=1\linewidth]{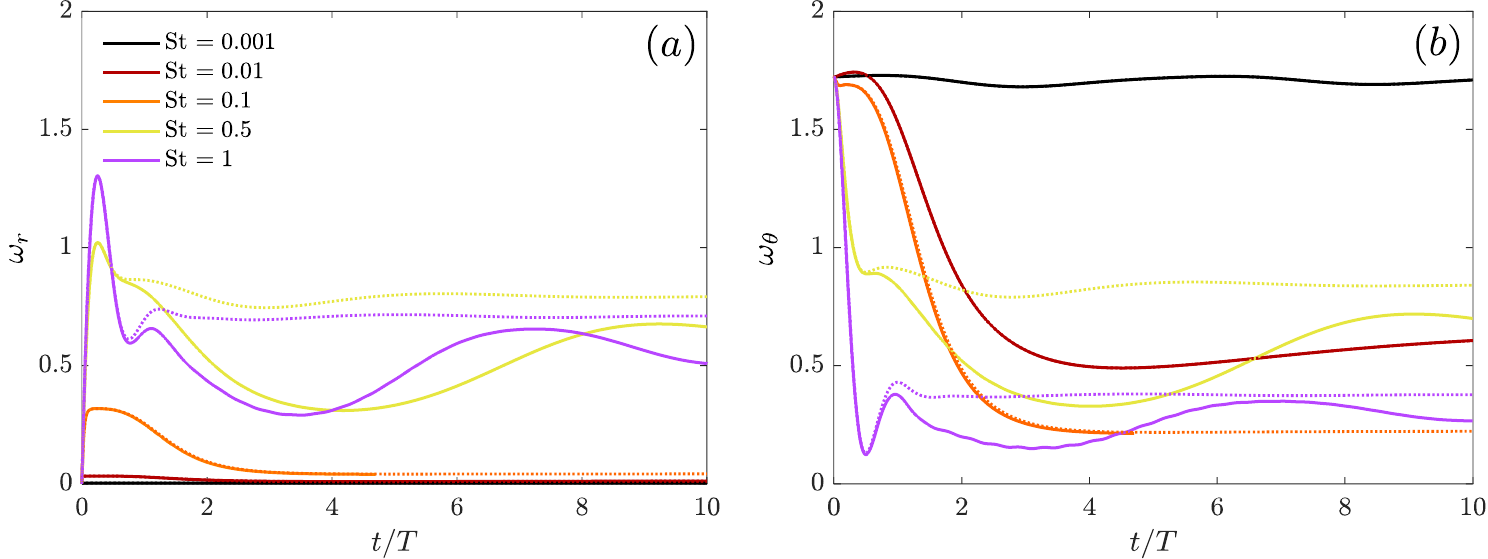}
    \caption{Rotational dynamics neutrally buoyant $\bar{\rho}=1$ spherical particles within a vortex rotating at a frequency $\phi_z = 2\pi$ for particle Stokes number in the range $\text{St} = 0.001 - 1$. Times are normalized by the vortex rotational period $T=\phi_z/2\pi$. (a) Particle radial axis rotation $\omega_r(t)$. (b) Particle tangential axis rotation of the particles $\omega_\theta(t)$.}
    \label{fig:3}
\end{figure}

\Cref{fig:3} presents the radial ($\omega_r$) and tangential ($\omega_\theta$) spin of neutrally buoyant spherical particles for different Stokes numbers, corresponding to the trajectories shown in \cref{fig:1,fig:2}.  
Radial axis spin initiates immediately across all Stokes numbers and is driven by the torque generated from the fluid's vorticity.
While increasing the Stokes number extends the initial response time, the overall behavior remains largely unaffected by variations in particle rotational inertia.  
In contrast, the tangential spin changes gradually as it is driven by particle translation, particularly in the intermediate Stokes range $0.01<\text{St}<0.1$.  
Although high-frequency oscillations have a minimal impact on the overall spin dynamics, slower Magnus-driven oscillations play a more prominent role. 
These low-frequency oscillations, resulting from substantial vertical displacements, influence both radial and tangential spins.  
Consequently, the emergence of these slow oscillations highlights the importance of Magnus lift forces in shaping particle spin, which, in turn, affects the linear particle motion.  
Thus, these results underscore the necessity of modeling rotational dynamics for particles of $\text{St} > 0.1$, even when the particles are perfectly spherical.

\begin{figure}
    \centering
    \includegraphics[width=1\linewidth]{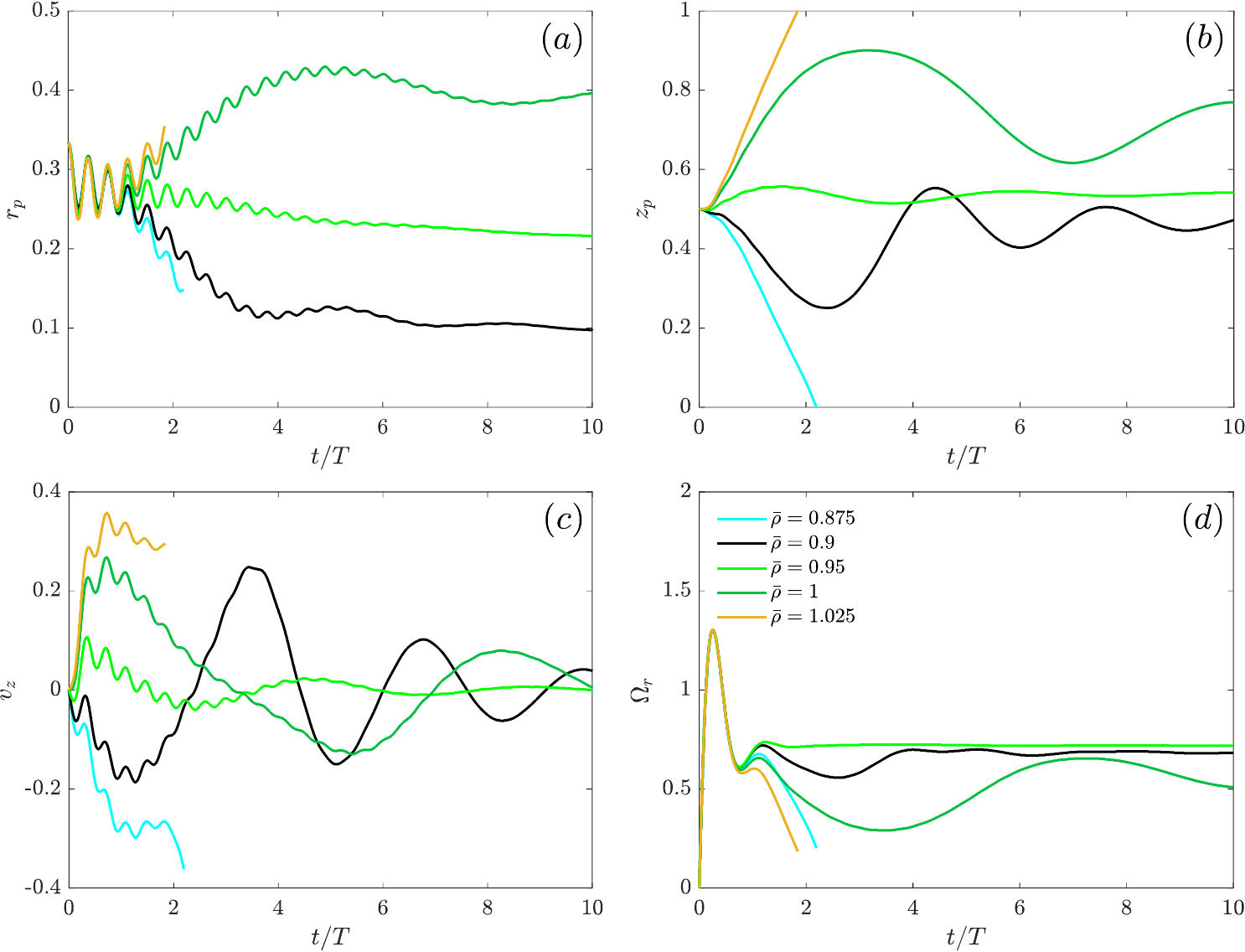}
    \caption{Dynamic response of $\text{St}=1$ particles within a vortex of constant frequency $\phi_z = 2\pi$. Particle-fluid density ratios vary in the range $\bar{\rho}=0.875 - 1.015$. Times are normalized by the vortex rotational period $T=\phi_z/2\pi$. (a) Particle radial location $r_p(t)$. (b) Particle axial location $z_p(t)$. (c) Particle axial velocity $v_z(t)$. (d) Particle radial axis rotation $\omega_r(t)$.}
    \label{fig:5}
\end{figure}

We now proceed to investigate the effects of other physical parameters on the particle response: the fluid-to-particle density ratio $\bar{\rho}$ and the vortex circulation frequency $\phi_z$.
The density ratio $\bar{\rho}$ represents the ratio of fluid density to particle density, where lower values of $\bar{\rho}$ correspond to heavier particles.
\Cref{fig:5} illustrates the dynamic response of $\text{St} = 1$ particles within a vortex of constant frequency, $\phi_z = 2\pi$, as the particle-fluid density ratio, $\bar{\rho}$, varies in the range 0.875 to 1.015. 
As expected, the heaviest and lightest particles sink or float out of the vortex cylinder domain, respectively.
For $0.9\leq\bar{\rho}\leq 1$, although the particle is heavier than the fluid, it does not sink out of the vortex. 
In fact, the heavy particle stabilizes at a smaller orbit compared to the neutrally buoyant particle. All the particles are initially drawn towards the cylinder axis, as evident in \cref{fig:5}(a).
While the centrifugal force acts to eject the heavy particles out towards the cylinder edge, the centrifugal force is surpassed by the virtual mass force near the flow stagnation point. 
Once pulled inwards, the particles settle into a periodic trajectory closer to the neutral axis, non-intuitively: the combined effects of Coriolis and virtual mass forces stabilize the competing centrifugal and drag forces. 
In contrast, the density ratio does not alter the fast-frequency radial oscillations, as they are governed primarily by drag forces.
Slower oscillations along the vertical axis, on the other hand, become faster as buoyancy and gravity effects increase, reflecting a less damped system.  
\Cref{fig:5}(d) suggests that the initial rotational dynamics are similar for all particles and independent of the density ratio. 
However, as the system evolves, the dynamics become irregular due to the particle spin and location coupling.

\begin{figure}
    \centering
    \includegraphics[width=1\linewidth]{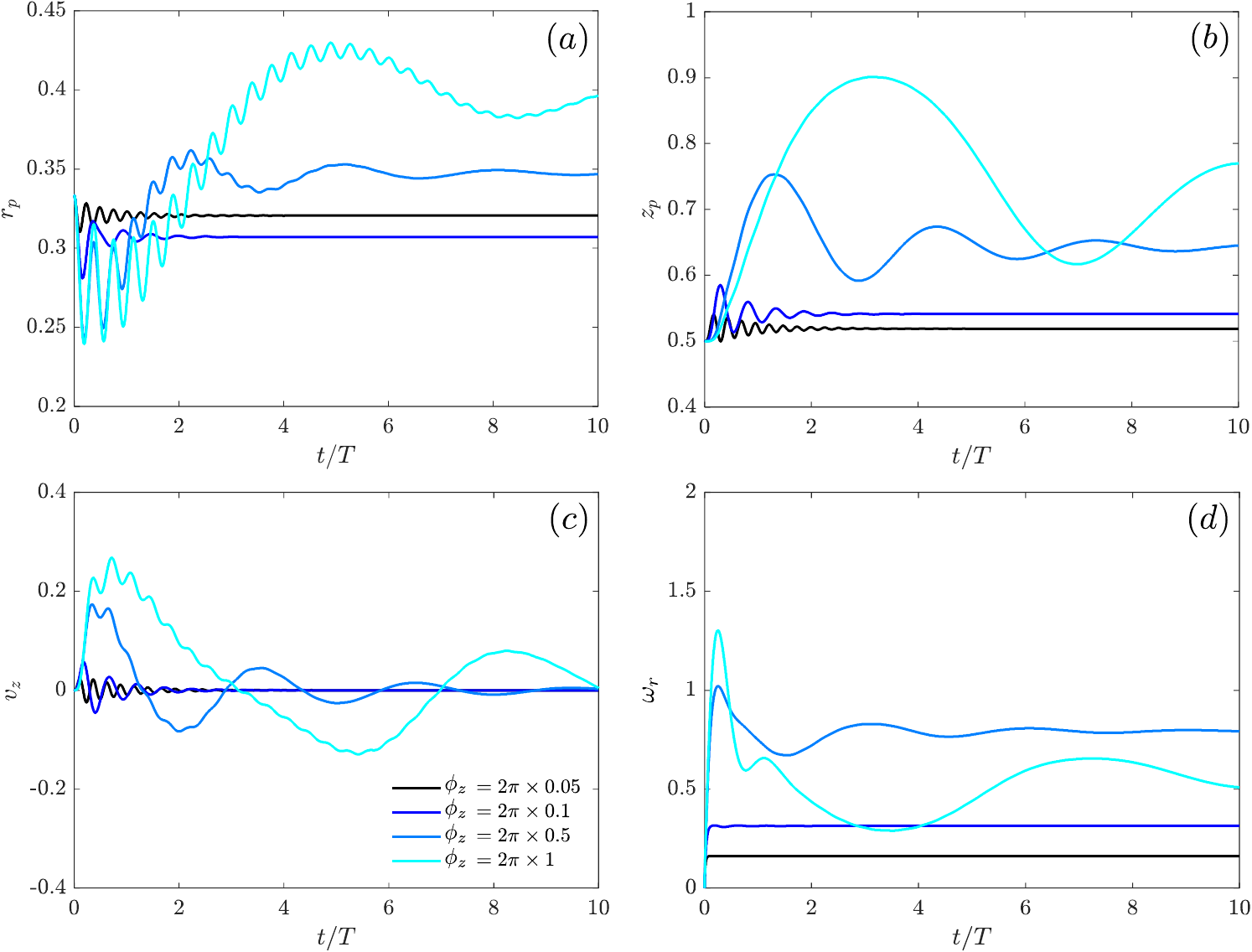}
    \caption{Dynamic response of $\text{St}=1$, neutrally-buoyant $\bar{\rho}=1$ spherical particles within a vortex of varying frequency $\phi_z = 2\pi\times(0.05 -1)$. Times are normalized by the vortex rotational period $T=\phi_z/2\pi$. (a) Particle radial location $r_p(t)$. (b) Particle axial location $z_p(t)$. (c) Particle axial velocity $v_z(t)$. (d) Particle radial axis rotation $\omega_r(t)$.}
    \label{fig:6}
\end{figure}

The vortex circulation frequency $\phi_z$ denotes the ratio between the vertical recirculation time and the azimuthal circulation time of the vortex flow.
Specifically, when $\phi_z<2\pi$, a fluid particle is projected to complete its vertical recirculation before finishing a full azimuthal orbit, whereas for $\phi_z>2\pi$, the azimuthal circulation occurs more rapidly than the vertical recirculation.
\Cref{fig:6} shows the dynamic response of Stokes number $\text{St}=1$, neutrally-buoyant spherical particles in a vortex with varying rotational frequencies, $\phi_z = 2\pi\times(0.05 - 1)$. The transient particle dynamics are plotted against the normalized time $t/T$, where $T = \phi_z/2\pi$ represents the vortex rotational period.
At lower rotational frequencies, virtual mass effects initially drive the particle orbit inward toward the neutral axis.
As seen in \cref{fig:6}(a), increasing $\phi_z$ beyond $2\pi\times0.5$ causes centrifugal forces to dominate, pushing the particles outward.
Simultaneously, the stable orbit is translated upwards as the rotation rate increases.
When normalized by $T$, we observed that the drag forces dissipate radial oscillations more rapidly at lower rotational frequencies. (rephrase)
On the other hand, radial and axial oscillations amplify with an increase in the rotational frequency; the vertical oscillation frequency decreases, thus becoming more energetic and decaying slower. 
This amplification suggests that a further increase in the rotational frequency could excite the particle to such an extent that it will be ejected out of the vortex, as it cannot reach a steady-state orbit.

\section{Steady State Orbits}\label{sec:ss} 
The dynamic patterns discussed thus far reveal that some particles settle into steady-state periodic orbits; these stable orbits exhibit sensitivity to both the vortex and particle properties. 
The emergence of such steady-state behavior manifests in trajectories of the form $\bm{x}(t)=(r_0,\theta(t),z_0)$, wherein the radial and axial coordinates remain fixed while the azimuthal position varies periodically, necessitating the following kinematic conditions for radial and axial velocities:
\begin{equation}\label{eq:cond1_eq}
v_{r,0}= v_{z,0} = 0,
\end{equation}
while the rest of the state variables maintain
\begin{equation}\label{eq:cond2_eq}
    \frac{dv_{\theta,0}}{dt} = \frac{d\omega_{r,0}}{dt} = \frac{d\omega_{\theta,0}}{dt} = \frac{d\omega_{z,0}}{dt} = 0.
\end{equation}
Upon substitution of these equilibrium conditions into \cref{eq:mom,eq:spin}, we derive a system of coupled nonlinear algebraic equations that govern the steady-state orbit,
\begin{equation}\label{eq:eq_state}
    \begin{gathered}
    -\frac{10}{3\text{St}}\omega_{r,0} + \phi_z\omega_{\theta,0} =0 \\
    \frac{10}{3\text{St}}(\Phi_\theta - \omega_{\theta,0}) - \phi_z\omega_{r,0} =0\\
    \frac{2U_0}{\text{St}} + 3\bar{\rho}\left[\left(\bm{u}_0\cdot\nabla\right)\bm{u}_0\right]_{r} - 2(2\bar{\rho} + 1)\phi_z^2r_0 + 2(2 + \rho)\phi_zv_{\theta,0} + \frac{3\bar{\rho}}{2}W_0(\Phi_\theta - \omega_{\theta,0}) =0\\
    \frac{v_{\theta,0} - \phi_z r_0}{\text{St}} + 3\bar{\rho} \left(U_0\phi_z  + \frac{1}{4}W_0\omega_{r,0}\right) = 0\\
    \frac{2W_0}{\text{St}} + 3\bar{\rho} \left[\left(\bm{u}_0\cdot\nabla\right)\bm{u}_0\right]_{z} - 2(1-\bar{\rho})Fr^{-2} - \frac{3\bar{\rho}}{2}\left[(\phi_z r_0 - v_{\theta,0})\omega_{r,0} + U_0(\Phi_\theta - \omega_{\theta,0})\right] =0.\\
\end{gathered}
\end{equation}
The local flow properties at the equilibrium position, denoted by $\Phi_\theta = \phi_\theta (r_0,z_0)$, $W_0 = u_z(r_0,z_0)$, and $U_0 = u_r(r_0,z_0)$, together with the convective derivatives at the equilibrium $\left(\bm{u}_0\cdot\nabla\right)\bm{u}_0$ and the dimensionless parameters $[\text{St},\bar{\rho},\phi_z]$, constitute the complete set of variables from which the steady-state solution $[r_0,z_0,v_{\theta,0},\omega_{r,0},\omega_{\theta,0}]$ can be resolved numerically.
Once the steady-state solution is obtained, we linearize the ODE system to extract the eigenvalues and characterize the equilibrium points.

\begin{figure}
    \centering
   \includegraphics[width=1\linewidth]{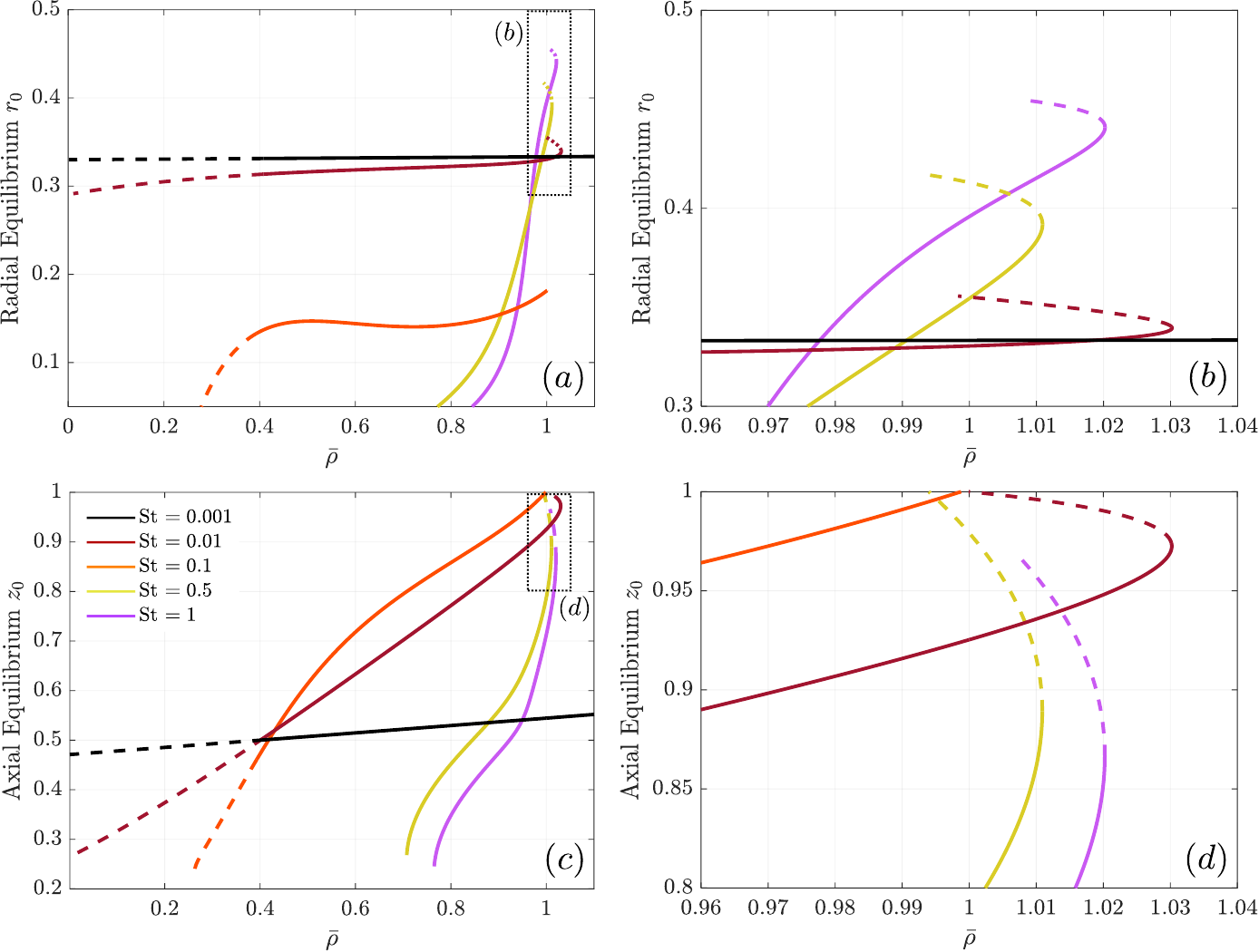}
    \caption{Particle equilibrium points as a function of its density ratio $\bar{\rho}$ for different particle $\text{St}$ and constant rotational frequency $\phi_z =2\pi$. Stable equilibria are denoted by a solid line while unstable ones are denoted by a dashed line. (a) Radial location of the equilibrium points. (b) Zoomed-in view of the bifurcation in the radial location of the equilibrium. (c) Axial location of the equilibrium points. (d) Zoomed-in view of the bifurcation in the axial location of the equilibrium.} 
    \label{fig:7}
\end{figure}

To uncover the effect of varying the density ratio on the equilibrium points, we vary the density ratio from $\bar{\rho}=0.001$, corresponding to solid particles in gaseous media, up to positively buoyant particles.
The nullclines presented in \cref{fig:7} allow one to observe how the particle density shifts the equilibrium point spatially; solid lines denote the stable equilibria, while dashed lines denote the unstable ones.
\Cref{fig:7} reveals that as particle size increases and drag becomes more influential, the minimal density ratio required for the existence of equilibrium  -- of any kind -- increases, as heavy inertial particles tend to sink out of the vortex.
Below a density ratio of $\bar{\rho} = 0.4$, the particles are too heavy to maintain a stable orbit, even for the smallest particles $\text{St} = 0.001$ -- where a tracer-like particle dynamics is expected; inertial effects prevent the particles from settling at a stable orbit and will eventually lead to them swirling away from the vortex. 

At the other end of the spectrum, low Stokes particles $\text{St} = 0.001$ stand out. 
Their equilibria stay close to the vortex center and retain stability even at positive buoyancy, consistent with their tracer-like dynamics. However, even a slight size increase to St$=0.01$ changes the orbit behavior: the axial equilibrium location translates upwards towards the cylinder top plane with decreasing particle density. 
This upward drift continues until the equilibrium vanishes beyond a critical density ratio, obtained here at $\bar{\rho}>1.03$.
This limit point behavior is shared across $\text{St} > 0.01$ particles; the introduction of Magnus forces in conjunction with even slight drag effects leads to buoyant particles being inherently unstable.
One may also observe that around the buoyancy-neutrality transition, a secondary branch of unstable equilibrium appears alongside the stable branch. 
As the density increases further, these branches converge at a saddle-node bifurcation, beyond which no equilibrium is sustained.

\begin{figure}
    \centering
   \includegraphics[width=\linewidth]{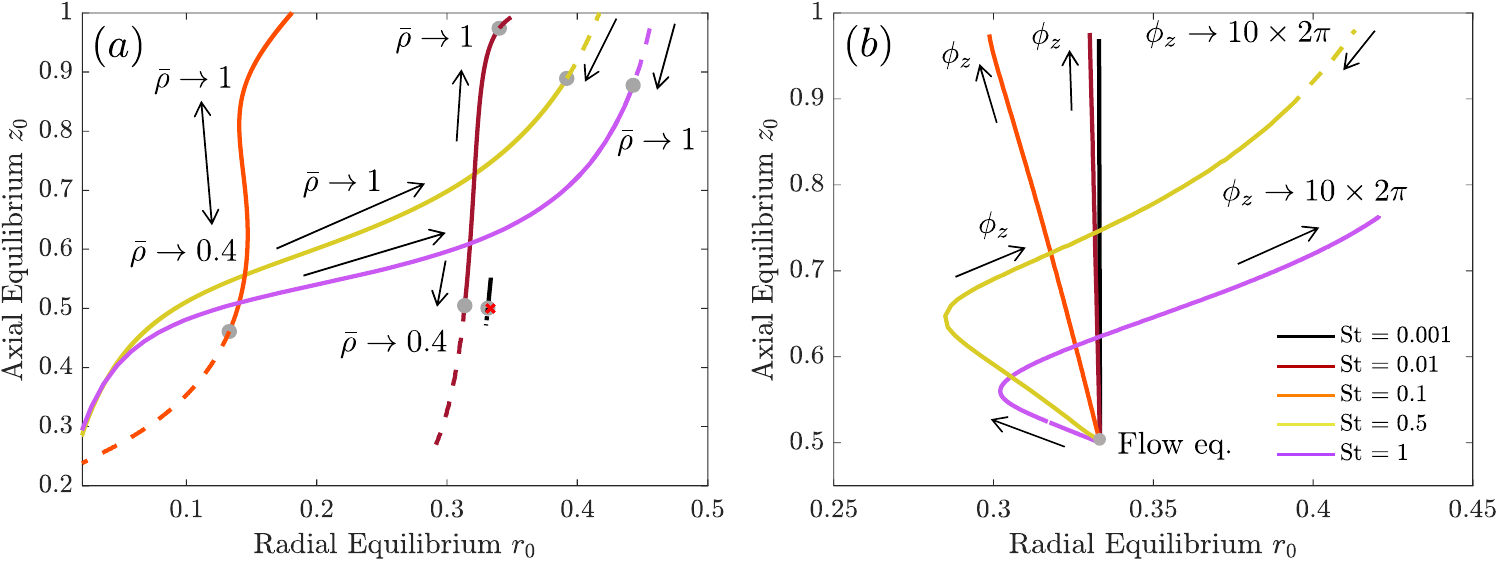}
    \caption{Spatial translation of the particle equilibrium points $(r_0, z_0)$ for varying Stokes numbers $\text{St} = 0.001 - 1$. Solid lines denote the stable equilibria, while dashed lines denote the unstable ones. (a) Spatial translation due to a continuous change of the particle density ratio $\bar{\rho}$ within a vortex field rotating at $\phi_z = 2\pi$. The transition between unstable and stable equilibrium is denoted by grey circles and the flow stagnation point by a red cross. (b) Spatial translation due to a continuous change of the vortex rotational frequency from $\phi_z=0$ to $\phi_z=2\pi\times10$ for a neutrally-buoyant particle $\bar{\rho}=1$.}
    \label{fig:8}
\end{figure}

\Cref{fig:8}(a) corresponds to the results of \cref{fig:7}, illustrating how equilibrium positions of the particles shift as the density ratio increases. 
One may observe the gradual effect of the increase in particle Stokes number; the equilibrium point of small particles remains close to that of fluid tracers, deviating only slightly from the flow equilibrium. 
Nevertheless, even relatively small inertial effects lead to deviations from purely tracer-like behavior, reinforcing the finding that weak drag effects could alter the particle's clustering pattern within the vortex.
As the particle inertia increases, the equilibrium translation exhibits a diagonal displacement across almost the entire cylindrical region of the vortex. 
This result suggests that higher-Stokes-number particles can maintain equilibrium even in regions of higher velocity and vorticity, whereas lower-inertia particles cannot. 
Similarly, \cref{fig:8}(b) examines the effect of increasing vortex rotational frequency on the equilibrium position of a neutrally buoyant particle. 
Initially, an increase in circulation causes the equilibrium position to shift inward and upward from the flow equilibrium; the radial translation intensifies for larger particle Stokes values.
This trend is consistent with earlier observations that virtual mass effects drive particles toward the neutral axis at moderate circulation rates. However, for sufficiently large particles, a reversal occurs -- beyond a critical circulation strength, the equilibrium position moves outward, dominated by increased centrifugal forces.

\section{Conclusion}\label{sec:conc}
This study examined the dispersion of spherical particles in an analytically defined, rotating cylindrical vortex -- a simplified model aimed at isolating the dominant physical mechanisms governing the transport of inertial and heavy particles in rotational flows.
To this end, the coupled translational and rotational equations of motion, incorporating drag, buoyancy, virtual mass, Coriolis, and Magnus lift forces, are derived.
The coupled equations set was numerically integrated over a broad range of particle Stokes numbers, density ratios, and vortex rotation frequencies.

The results demonstrate that particle aggregation and orbital stability stem from the complex coupling between all of the modeled forces.
Rotational lift forces, often overlooked when considering spherical particles, were found to be negligible at small Stokes numbers but become important for $\text{St} \sim 1$, inducing slow vertical oscillations in both position and particle spin.
The coupling between spin and translational motion further affects orbital stability and determines the asymptotic states reached by the particles.
The fluid–particle density ratio $\bar{\rho}$ modulates the force balance; heavier particles ($\bar{\rho} < 1$) can remain trapped in the vortex under specific conditions, forming periodic orbits stabilized by Coriolis and virtual mass effects, while lighter particles ($\bar{\rho} > 1$) exhibit enhanced vertical oscillations and are more likely to escape.

Under certain conditions, particles settle into steady-state orbits governed by a nonlinear equilibrium system. These equilibria depend non-monotonically on particle and flow parameters, with bifurcations giving rise to multiple or vanishing solutions.
Such behavior helps identify particle classes prone to clustering due to dominant inertial effects in vortical flows, with potential applications in marine ecology and industrial separation processes.
Future extensions of this framework may incorporate non-spherical particles, oscillatory flow fields, and mass biofueling processes. 

\newpage
\bibliography{references}

\end{document}